\UseRawInputEncoding 
\documentclass[reprint,superscriptaddress,amsmath,amssym,prb]{revtex4-2}

\usepackage{bm}
\usepackage{float}
\usepackage{graphicx}
\usepackage[usenames,dvipsnames]{color}
\usepackage[normalem]{ulem}
\usepackage[svgnames]{xcolor}
\usepackage{bm}% bold math
\usepackage{multirow}
\usepackage{float}
\usepackage{titlesec}
\usepackage[utf8]{inputenc}

\usepackage{color}

\begin{document}
%\preprint{APS/123-QED}

\title{Glass-like thermal conductivity and narrow insulating gap of EuTiO$_3$}

\author{Alexandre Jaoui}\thanks{Equal contribution\\ Present address: Fakult\"at f\"ur Physik, Ludwig-Maximilians Universit\"at München, Geschwister-Scholl-Platz 1, 80539 M\"unchen, Germany}
\affiliation{JEIP, USR 3573 CNRS, Coll\`ege de France, PSL Research University, 11, Place Marcelin Berthelot, 75231 Paris Cedex 05, France}\affiliation{Laboratoire de Physique et d'\'Etude des Mat\'eriaux \\ (ESPCI - CNRS - Sorbonne Universit\'e), PSL Research University, 75005 Paris, France}

\author{Shan Jiang}\thanks{Equal contribution}
\affiliation{Laboratoire de Physique et d'\'Etude des Mat\'eriaux \\ (ESPCI - CNRS - Sorbonne Universit\'e), PSL Research University, 75005 Paris, France}

\author{Xiaokang Li}\thanks {Present address: Wuhan National High Magnetic Field Center and School of Physics, Huazhong University of Science and Technology, Wuhan 430074, China}
\affiliation{Laboratoire de Physique et d'\'Etude des Mat\'eriaux \\ (ESPCI - CNRS - Sorbonne Universit\'e), PSL Research University, 75005 Paris, France}

\author{Yasuhide Tomioka}\affiliation{National Institute of Advanced Industrial Science and Technology (AIST), Tsukuba 305-8565, Japan}

\author{Isao H. Inoue}\affiliation{National Institute of Advanced Industrial Science and Technology (AIST), Tsukuba 305-8565, Japan}

\author{Johannes Engelmayer}
\affiliation{II. Physikalisches Institut, Universit\"at zu K\"oln, 50937 K\"oln, Germany}

\author{Rohit Sharma}
\affiliation{II. Physikalisches Institut, Universit\"at zu K\"oln, 50937 K\"oln, Germany}

\author{Lara P\"atzold}
\affiliation{II. Physikalisches Institut, Universit\"at zu K\"oln, 50937 K\"oln, Germany}

\author{Thomas Lorenz}
\affiliation{II. Physikalisches Institut, Universit\"at zu K\"oln, 50937 K\"oln, Germany}

\author{Benoît Fauqu\'e}%
\affiliation{JEIP, USR 3573 CNRS, Coll\`ege de France, PSL Research University, 11, Place Marcelin Berthelot, 75231 Paris Cedex 05, France}

\author{Kamran Behnia}
\affiliation{Laboratoire de Physique et d'\'Etude des Mat\'eriaux \\ (ESPCI - CNRS - Sorbonne Universit\'e), PSL Research University, 75005 Paris, France}

\date{\today}

\begin{abstract}
Crystals and glasses differ by the amplitude and the temperature dependence of their thermal conductivity. However, there are crystals known to display glass-like thermal conductivity. Here, we show that EuTiO$_3$, a quantum paraelectric known to order antiferromagnetically at 5.5 K, is one such system. The temperature dependence of resistivity and Seebeck coefficient yield an insulating band gap of $\sim 0.22$ eV. Thermal conductivity is drastically reduced. Its amplitude and temperature dependence are akin to what is seen in amorphous silica. Comparison with non-magnetic perovskite solids, SrTiO$_3$, KTaO$_3$, and EuCoO$_3$, shows that what impedes heat transport are $4f$ spins at Eu$^{2+}$ sites, which couple to phonons well above the ordering temperature. Thus, in this case, superexchange and valence fluctuations, not magnetic frustration, are the drivers of the glass-like thermal conductivity.
\end{abstract}
\maketitle

\section{Introduction}
In most insulating crystals, the flow of heat can be understood by considering the response of a gas of phonons to a temperature gradient. This picture, first drawn by Peierls \cite{Peierls1929}, using a linearized Boltzmann equation, is remarkably successful in describing the thermal conductivity, $\kappa$, of most insulators \cite{Lindsay2013,subedi21}. It explains why at an intermediate temperature, $\kappa$ peaks \cite{vandersande1986}, separating a high-temperature decrease by warming (due to anharmonicity), and a low-temperature decrease by cooling (due to phonon depopulation). In amorphous solids, on the other hand, there is no such  peak in $\kappa (T)$ and heat diffuses thanks to off-diagonal coupling across harmonic branches \cite{Allen1999,seyf2016,Simoncelli2019,Isaeva2019}. 

Some crystals, however, display glass-like thermal conductivity \cite{Visser1997,Takahata2000,Li2013,TACHIBANA201316,Sugii2017,Dutta2019,HANUS2021100344}. The thermal conductivity of these materials can be very low and/or feature a monotonic temperature dependence (lacking the $T^{-1}$ decrease at high temperature). They are sought after, since a low lattice thermal conductivity would lead to a large thermoelectric figure of merit \cite{Snyder2008}.

EuTiO$_3$ (ETO) is a perovskite with a cubic structure at room temperature \cite{Bussmann_Holder_2012}. Its electric permittivity increases upon cooling without giving rise to a ferroelectric instability. This quantum paraelectric behavior \cite{Muller1979} is akin to what has been seen in other ABO$_3$ compounds such as SrTiO$_3$ (STO) and KTaO$_3$ (KTO). In contrast to them, however, it magnetically orders at T$_N=5.5$ K, with an antiferromagnetic alignment of the nearest neighbor Eu$^{2+}$ spins \cite{McGuire,Katsufuji2001}. Like STO, but at higher temperatures, it goes also through an antiferrodistortive transition where adjacent TiO$_6$ octahedra rotate in opposite directions \cite{Bussmann2011}. Upon doping, either with 
oxygen vacancies \cite{Engelmeyer2019} or La substitution of Ti \cite{Tomioka2018}, it becomes a dilute metal \cite{Engelmeyer2019,Tomioka2018,Maruhashi2020}, but without a superconducting ground state.   

Here, we show that the temperature dependence of thermal conductivity in ETO is glass-like and its amplitude, over a broad temperature range starting from room temperature and extending to cryogenic temperatures, is lower than in STO and KTO. The drastic attenuation of thermal conductivity occurs well above the N\'eel temperature and when the magnetic entropy is saturated to its expected value. This points to an unusual version of spin-phonon coupling such as the phonon-paramagnon hybridization postulated by Bussmann-Holder \textit{et al.} \cite{Bussmann_Holder_2012}. 

Evidence for spin-lattice coupling in ETO has been around for more than two decades \cite{Katsufuji2001}. However, it was restricted to temperatures comparable to the magnetic ordering. According to our findings, the phonon mean free path is affected by spins at Eu sites even at high temperature when there is neither a magnetic order nor a field-dependent entropy. 

Monitoring the temperature dependence of electrical conductivity and thermopower in insulating EuTiO$_3$, we find an activation gap of 0.11 eV, indicating that the intrinsic transport band gap in EuTiO$_3$ is as low as $\sim$0.2 eV, much smaller than the gaps found by optical studies, but close to what is expected by \textit{Ab Initio} calculations \cite{Ranjan_2007,Malyi2022}. 

The narrow gap between the chemical potential and Eu$^{2+}$ energy level and the random orientation of large magnetic moments in the paramagnetic state emerge as principal suspects shortening the lifetime of heat carrying phonons at elevated temperatures in the paramagnetic state.  

In contrast to many other crystals displaying glass-like conductivity, ETO is not a spin liquid candidate, but a simple $ G$-type antiferromagnet and there is no magnetic frustration. In this context, the glassy behavior could be framed in a picture of off-diagonal coupling  across harmonic modes \cite{Simoncelli2019,HANUS2021100344}, which in this case would be phonons and paramagnons. 

\begin{figure}
\begin{center}
\centering
\makebox{\includegraphics[width=0.45\textwidth]{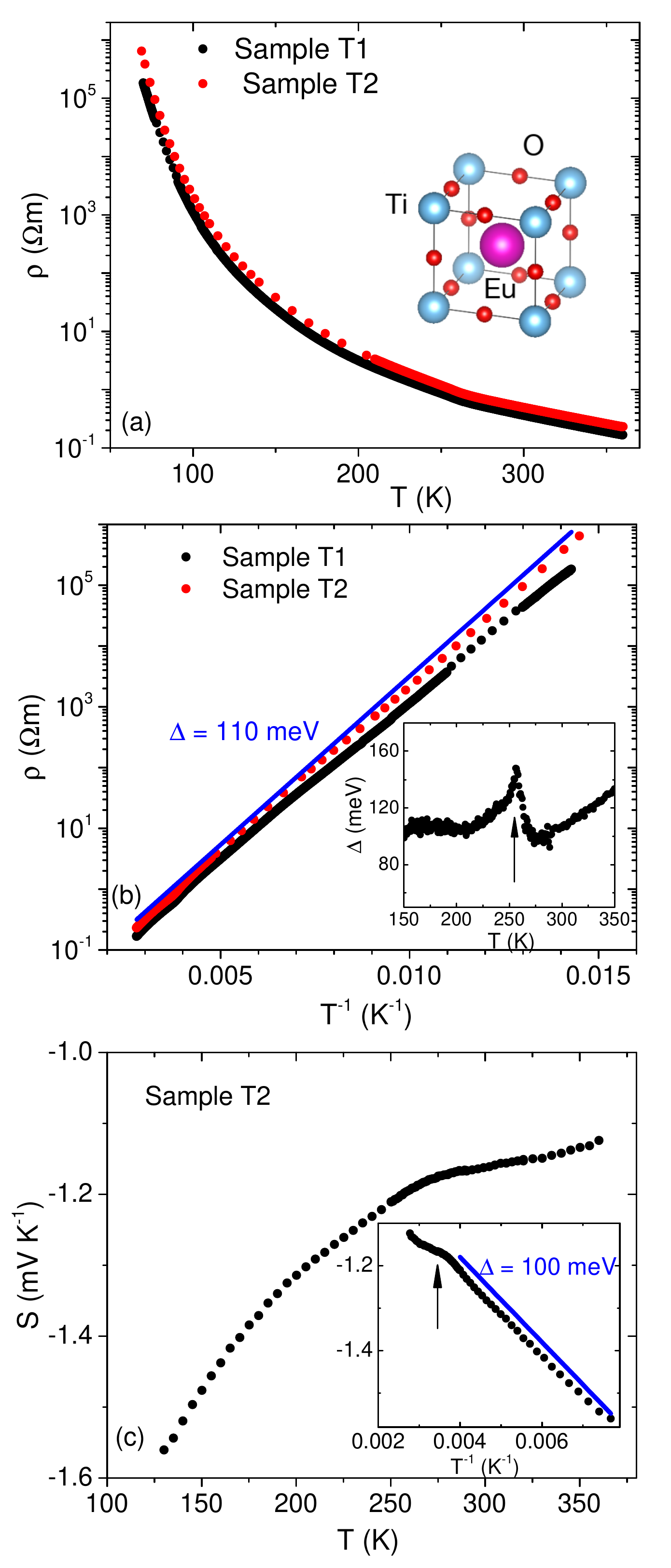}}
\caption{{\bf{Electrical resistivity, thermopower and activation gap of EuTiO$_3$:} } a) Resistivity, $\rho$, as a function of temperature in two single crystals of EuTiO$_3$ in a semi-logarithmic plot. The inset shows the unit cell (in the cubic phase). b) Arrhenius plot of the same data: $ln \rho$ \textit{vs.}  $T^{-1}$. The solid blue line corresponds to $\Delta=0.11eV$. The inset shows the temperature variation of $\Delta=-k_BT^2\frac{\partial ln\rho}{\partial T}$, with an arrow pointing to $T_{AFD}$, the temperature of antiferrodistortive transition. c) Seebeck coefficient, $S$ as a function of temperature. The data is restricted to $T>120 K$, below which measurement becomes problematic. The inset shows $S$ as a function of $T^{-1}$. The solid blue lines corresponds to $\Delta=0.1eV$. The arrow points to $T_{AFD}$.}
\label{rho}
\end{center}
\end{figure}

\section{Results}
\subsection{Activation and band gap}
Fig.\ref{rho}a shows the temperature dependence of electrical resistivity $\rho$ in as-grown EuTiO$_3$ (ETO) single crystals. One can see that $\rho$ increases by seven orders of magnitude upon cooling from 360 K to 50 K. 

An Arrhenius activation behavior becomes visible by plotting ln$\rho$ as a function of the inverse of temperature (Fig.\ref{rho}b). The activation gap is $\sim 0.11$ eV. 
The Hall resistivity (see the supplement \cite{SM}) shows also an Arrhenius behavior with a comparable $\Delta$.  At low temperature, resistivity begins to saturate when the Hall carrier density is  as low as $10^{10} cm^{-3}$ (See the supplement \cite{SM}). This implies an extremely low level of extrinsic donors . An activated behavior in longitudinal and Hall resistivities of as-grown ETO crystals was previously reported by Engelmayer \textit{et al.} \cite{Engelmeyer2019}, whose study focused on the emergence of metallicity in oxygen-deficient ETO.  

This activated behavior is also confirmed by our measurements of the Seebeck coefficient (See Fig.\ref{rho}(c)) and and the temperature dependence of the electric permittivity \cite{Engelmeyer2019}.

The thermoelectric power has a negative sign and an amplitude in the range of mV/K, typical of a narrow gap semiconductor. The Seebeck coefficient in an intrinsic semiconductor is expected to be $\approx\frac{k_B}{e}\frac{\Delta}{k_BT}$ \cite{Behnia2015b}. Thus, by plotting $S$ as a function of $T^{-1}$ (see the inset), one expects to see a straight line whose slope yields $\Delta$. As seen in the inset of Fig.\ref{rho}c, this is indeed what our data yields, with $\Delta \approx 0.1 eV$. Note that this temperature dependence is qualitatively distinct from what is expected in extrinsic semiconductors \cite{Johnson1953} as seen, for example, in the case of Nb-doped STO \cite{Collignon2020}. 

Thus, the temperature dependence of electric and thermoelectric conductivity both point to a similar energy gap between the chemical potential and the conduction band. The Hall and the Seebeck coefficients are both negative, indicating that carriers are electron-like and thermally excited to live in the conduction band originating from Ti$-d$ orbitals.

Both $\rho$ and $S$ show an anomaly near 260 K, which we identify as the temperature of the structural transition in ETO \cite{Bussmann_Holder_2012,Rushchanskii2012}. Like STO \cite{Lytle1964}, this transition makes ETO tetragonal \cite{Kennedy_2014}. We cannot rule out a very small difference in the amplitude of the activation gap between the cubic and the tetragonal phases (See the inset of Fig\ref{rho}b). We also detected  an unexpected and reproducible hysteresis near this phase transition. Taken at its face value, this indicates that this structural phase transition is first order (see the supplement).   

Assuming that the chemical potential is at halfway between the chemical potential and the valence band leads us to conclude that the band gap of ETO is $\approx 0.22$ eV. This is remarkably smaller than the 3.2 eV gap of STO \cite{Collignon2019}, but only slightly smaller than what a recent DFT calculation \cite{Malyi2022} found (0.27-0.33 eV). According to another and earlier theoretical study \cite{Ranjan_2007}, the magnitude of the band gap in ETO depends on the amplitude of the Hubbard $U$, and  a realistic $U$ (5-6 eV) would lead to a band gap of 0.2-0.4 eV. Thus, the narrow gap detected by our transport studies is close to what is intrinsically expected in this solid. Optical probes, however, have detected a larger gap (0.8-0.9 eV) \cite{Lee2009,Akamatsu2012}. Such larger energy scales are possible indications that the Density of States (DOS) near the Eu$-f$ level is not featureless \cite{Malyi2022}. 

Let us keep in mind the contrast between ETO on one hand and STO and KTO on the other hand. The first has a valence band originating from Eu$-f$ orbitals close to the chemical potential, while the two others have a valence Band emanating from O$-p$ orbitals and much further away from the chemical potential.

\subsection{Thermal conductivity}
\begin{figure}
\begin{center}
\centering
\makebox{\includegraphics[width=0.5\textwidth]{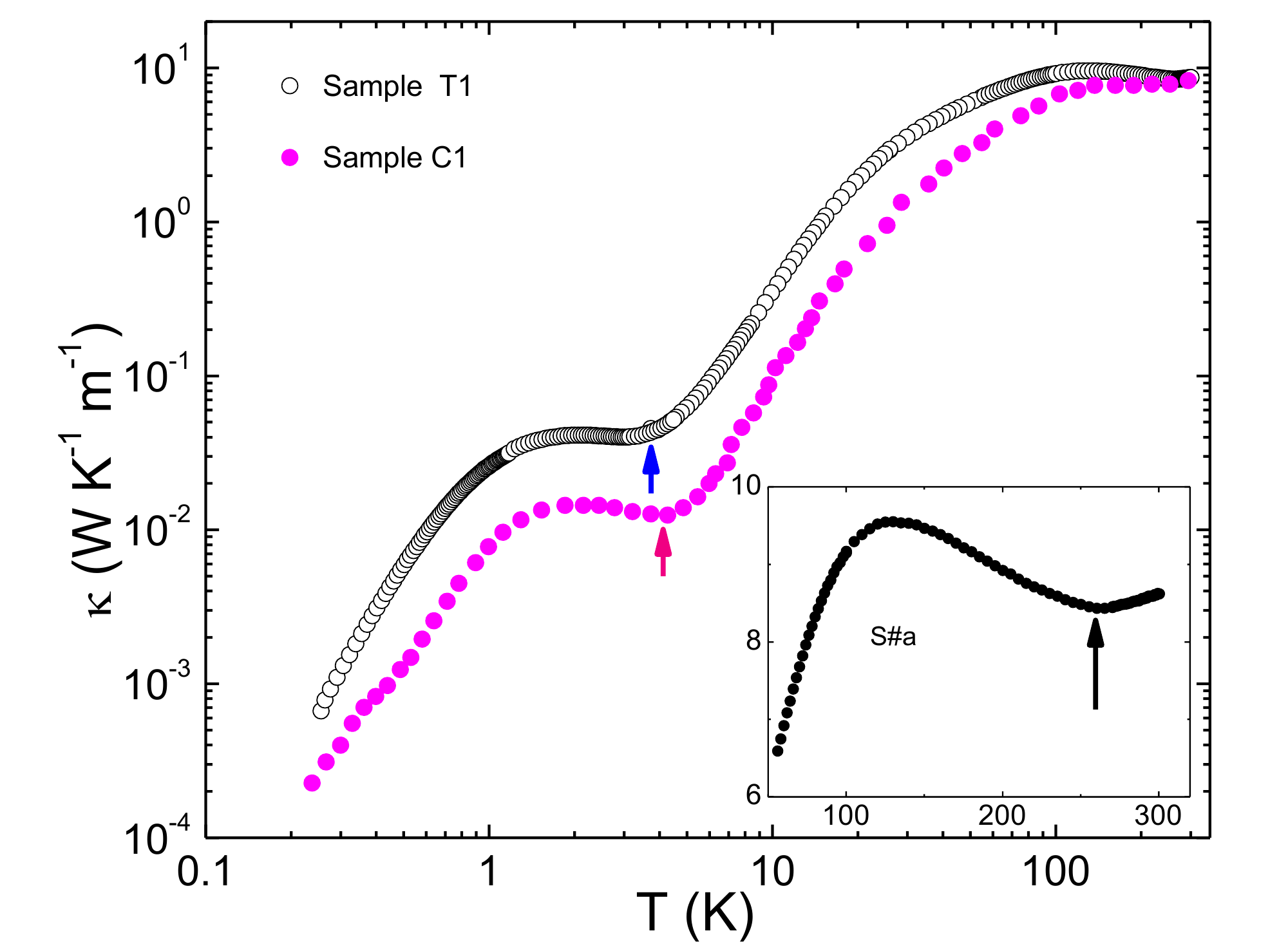}}
\caption{{\bf{Thermal conductivity of two EuTiO$_3$ crystals:} } a) Thermal conductivity, $\kappa$, as a function of temperature in two EuTiO$_3$ crystals in a log-log plot. The arrow points to a minimum near the N\'eel temperature. The inset shows a linear plot, with an arrow pointing to $T_{AFD}$.}
\label{kappa}
\end{center}
\end{figure}

Fig. \ref{kappa} shows the temperature dependence of thermal conductivity, $\kappa$, of two ETO crystals from slightly above room temperature down to dilution refrigerator temperatures. We measured several crystals and found that the thermal conductivity of all lies somewhere between the two cases shown in this figure. Moreover, we could not detect a correlation between the amplitude of $\kappa$ and a weak variation of the saturation magnetization observed across various ETO samples (see the supplement \cite{SM}).

In contrast to typical crystalline insulators \cite{vandersande1986}, $\kappa$ does not show a prominent peak. As seen in the inset,  there is a clear anomaly at $T_{AFD}$ and below this temperature $\kappa$ rises by ten percent increase with cooling. However, there is no sign of a kinetic regime with $\kappa \propto T^{-1}$, as seen in many other insulators \cite{Berman:1976,Xu2021}.

\subsection{Crystals, glasses and glass-like crystals}

\begin{figure}
\begin{center}
\centering
\makebox{\includegraphics[width=0.52\textwidth]{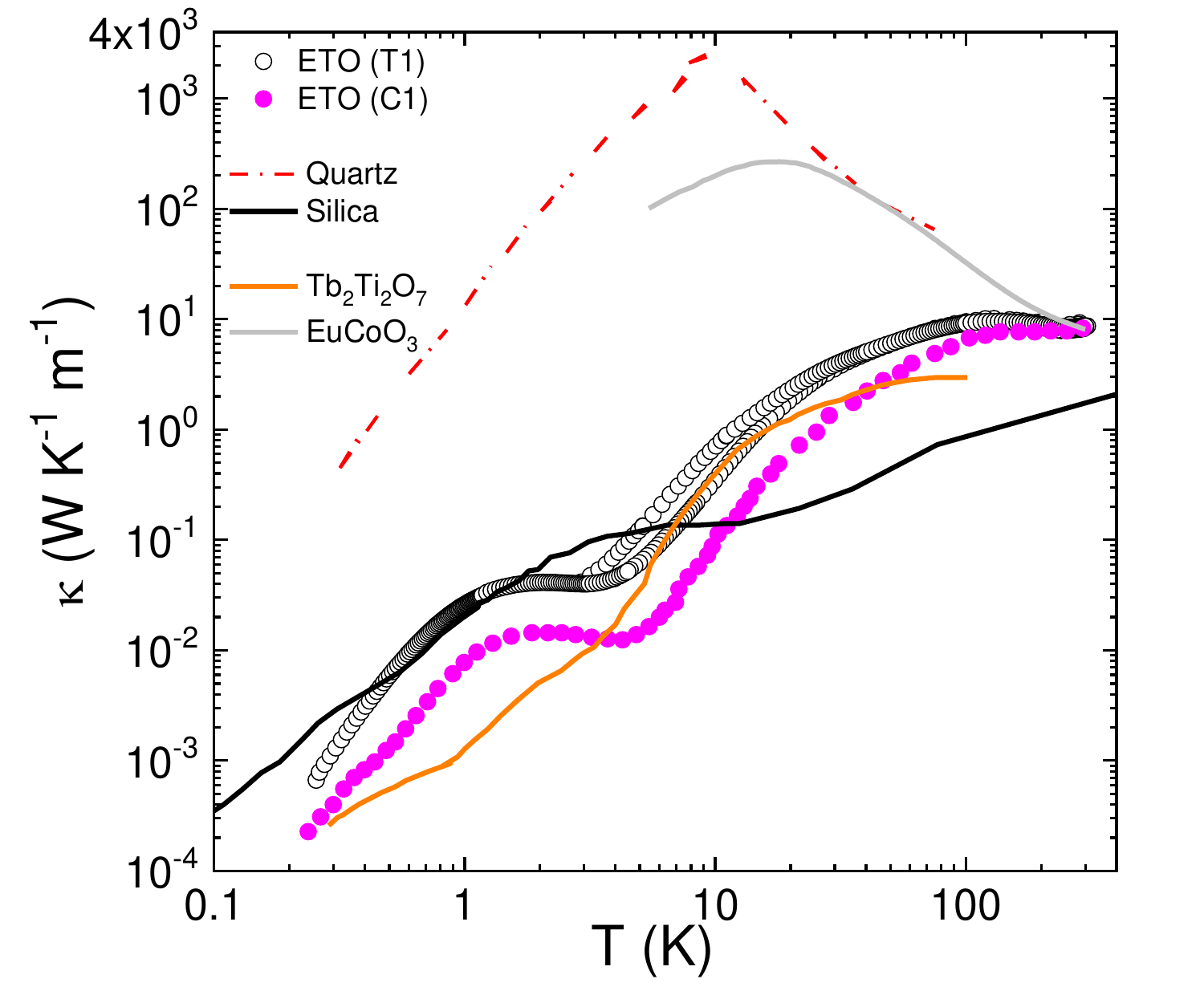}}
\caption{{\bf{Comparison with crystals, glasses and glass-like crystals:}} Thermal conductivity as a function of temperature in crystalline quartz, in vitreous silica \cite{Zeller1971}, in EuTiO$_3$ and in the frustrated magnet Tb$_2$Ti$_2$O$_7$ \cite{Li2013}. The latter two crystalline compounds conduct heat like a glass rather than like a crystal. Also shown is the crystal-like thermal conductivity of non-magnetic EuCoO$_3$~\cite{Berggold2008}.}
\label{compa-glasses}
\end{center}
\end{figure}

In order to put our observation in a proper context, we compare our data with what has been reported in the case of SiO$_2$, which shows a spectacular difference in the thermal conductivity in its crystalline and amorphous structures \cite{Zeller1971}. As seen in  Fig.\ref{compa-glasses}, $\kappa$ in amorphous silica monotonically decreases with cooling, in contrast to crystalline quartz, which has a prominent peak. At any given temperature, the crystal conducts heat at least an order of magnitude more than the glass \cite{vandersande1986}. Not only the thermal conductivity of ETO is similar to silica in temperature dependence, but also in the cryogenic temperature range surrounding the N\'eel temperature ($T_N\simeq 5.5 $K), the ETO crystalline samples conduct heat less than amorphous silica. 
The order of magnitude of thermal conductivity and its temperature dependence in ETO is comparable with other crystalline solids displaying a glass-like thermal conductivity, such as Tb$_2$Ti$_2$O$_7$ \cite{Li2013}, Tb$_3$Ga$_5$O$_{12}$ \cite{Inyushkin2010}, Na$_4$Ir$_3$O$_8$ \cite{Fauque2015}, Pr$_2$Ir$_2$O$_7$ \cite{Uehara2022} and La$_{0.2}$Nd$_{0.4}$Pb$_{0.4}$MnO$_4$ \cite{Visser1997}. In order to allow a direct comparison, Fig.\ref{compa-glasses} includes the data for Tb$_2$Ti$_2$O$_7$ \cite{Li2013}, the compound with the lowest thermal conductivity among these frustrated magnets. Note that in contrast to other members of this club of materials, EuTO$_3$ is a G-type antiferromagnet and is not frustrated.  

As a further comparison, we also include $\kappa$ of EuCoO$_3$~\cite{Berggold2008}, which displays a temperature dependence typical of a crystalline insulator. EuCoO$_3$ is a perovskite like ETO, but it has an orthorhombic symmetry, and the valence is Eu$^{3+}$ with only 6 electrons in the 4f shell. According to Hund's rules, this causes a vanishing magnetic moment ($J=L-S=0$), in agreement with the experimentally observed van Vleck susceptibility \cite{Baier2005}, drastically different from the large local moments of $7\,\mu_B/$Eu$^{2+}$ in ETO. 

\begin{figure}
\begin{center}
%\centering
\makebox{\includegraphics[width=0.5\textwidth]{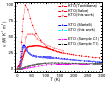}}
\caption{{\bf{Thermal conductivity of three quantum para-electric solids:}} Thermal conductivity of the two EuTiO$_3$ samples in a linear scale compared with thermal conductivity of  SrTiO$_3$ \cite{Martelli2018} and KTaO$_3$ \cite{Salce_1994,TACHIBANA201316}. Note the drastic reduction of thermal conductivity in the paramagnetic EuTiO$_3$. The visible sample dependence of the data for each material is much smaller than the differnce between the three compounds.}
\label{compa-QP}
\end{center}
\end{figure}

\subsection{ETO compared to its non-magnetic sisters}

Fig.\ref{compa-QP} compares the thermal conductivity of ETO and two other ABO$_3$ perovskites. SrTiO$_3$ (STO) and KTaO$_3$ (KTO) are also quantum paraelectric, but not magnetic, solids. Our new data on these two materials is in good agreement with previous studies of heat transport in STO \cite{Steigmeier1968,Martelli2018,Li2020} and in KTO \cite{Steigmeier1968,Salce_1994,Li2020}. In both cases, there is also a visible sample dependence, which is more pronounced near the peak. However, this sample dependence is much smaller than the difference between the three compounds. At room temperature, this difference is small, yet visible:  $\kappa (300 K)$ is $\approx 9$ W/(K.m) in ETO, $\approx 11$ W/(K.m) in STO and $\approx 17$ W/(K.m) in KTO. Much more drastic is the difference in the temperature dependence between the three sister compounds. The enhancement with cooling observed in the other perovskites is absent in ETO.  This difference extends over the full temperature range above the magnetic ordering. 

%\begin{figure}
%\begin{center}
%\centering
%\includegraphics[width=0.49\textwidth]{Figure/Fig2draft2.pdf}
%\caption{{\bf{Comparison to other solids:}} a) $\kappa$ vs. T in log-log scale for KTaO$_3$, SrTiO$_3$ and EuTiO$_3$. b) Comparison of $\kappa$ of EuTiO$_3$ with other  magnetic materials (Tb$_2$Ti$_2$O$_7$ \cite{Li2013}, Tb$_3$Ga$_5$O$_12$ \cite{Inyushkin2010}, Na$_4$Ir$_3$O$_8$ \cite{Fauque2015}, Pr$_2$Ir$_2$O$_7$ \cite{Uehara2022} and La$_{0.2}$Nd$_{0.4}$Pb$_{0.4}$MnO$_4$ \cite{Visser1997}) displaying as well a glassy like $\kappa$ behavior at low temperature.}
%\label{Fig2}
%\end{center}
%\end{figure}

\begin{figure}
\begin{center}
\centering
\includegraphics[width=0.49\textwidth]{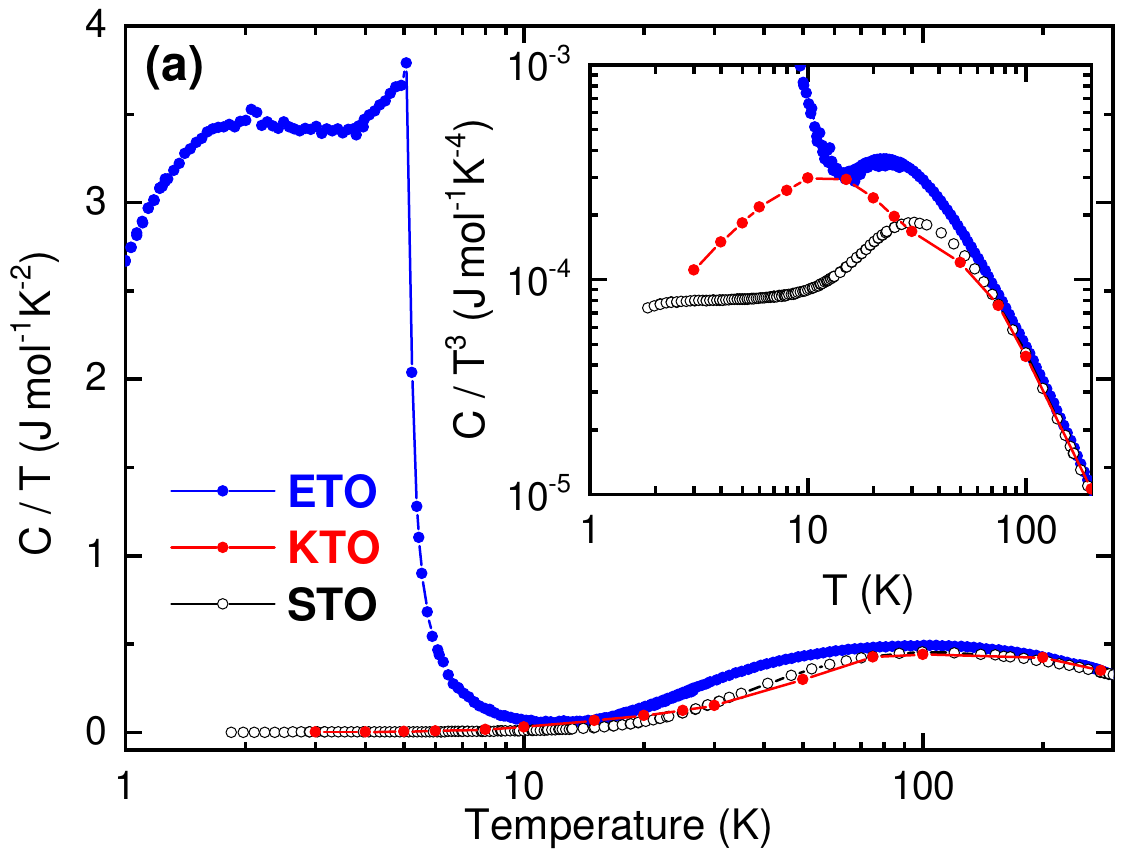} 
\includegraphics[width=0.49\textwidth]{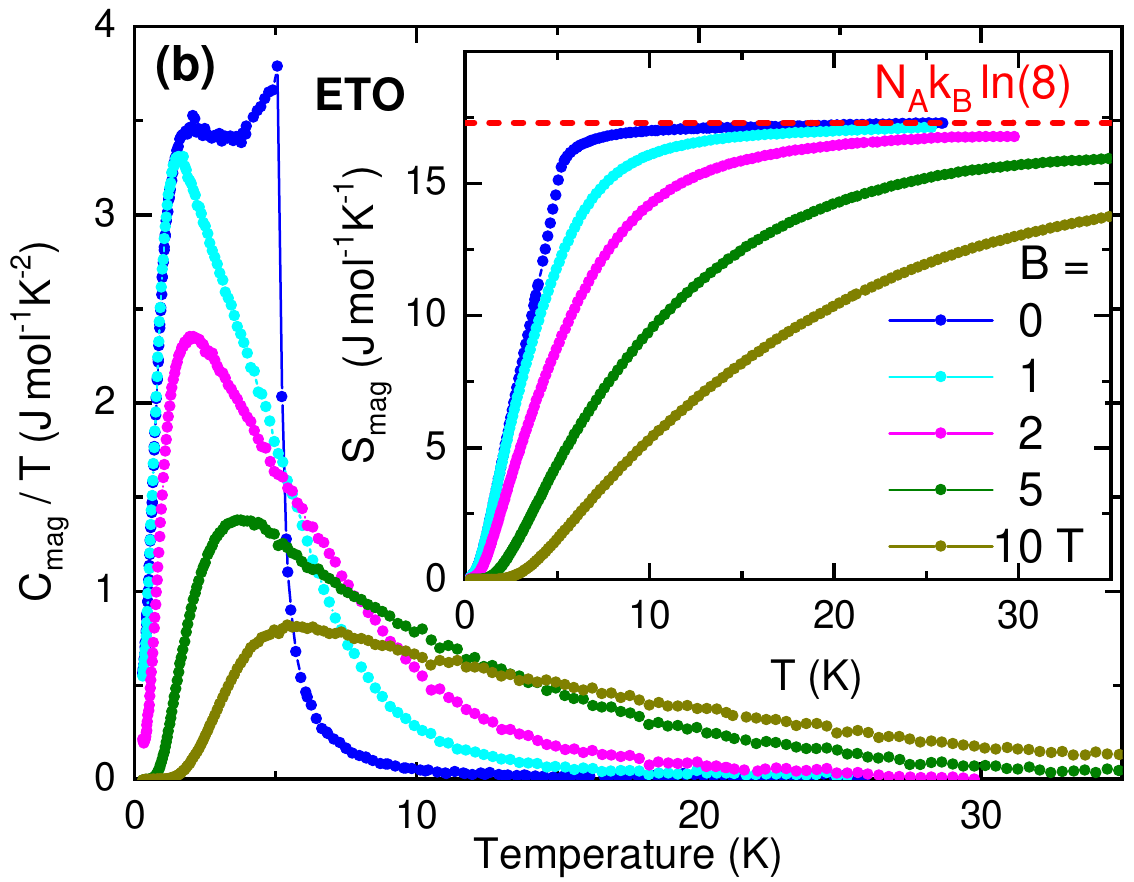}
\caption{{\bf{Specific heat in three quantum paraelectric solids:}} a) Temperature dependence of the total specific heat in KTaO$_3$, SrTiO$_3$ and EuTiO$_3$ plotted as $C/T$ \textit{vs.} $T$. The inset is an enlarged view of the same data shown as $\frac{C_p}{T^3}$ \textit{vs.} $T$. Note the presence of Einstein modes in all three solids and the additional magnetic specific heat in ETO emerging below 15 K. b) Evolution of $C_{mag}/T$ of ETO for different magnetic fields up to 10~T. The inset shows the corresponding  magnetic entropy $S_{mag} = \int C_{mag}/T\,dT$ of ETO and the maximum magnetic entropy $N_A k_B \ln(2S+1)$ of a spin 7/2 system is indicated by the dashed line.} 
\label{Cp_QP}
\end{center}
\end{figure}

It is instructive to scrutinize the specific heat of the three sister compounds. Fig.~\ref{Cp_QP}(a) shows that $C/T$ of STO, ETO and KTO approach each other towards room temperature and reach $C \approx 100\,$J/(mol.K). With five atoms, one expects the specific heat saturating to the Dulong-Petit value of $5\,N_A k_B=125\,$J/(mol.K), which is indeed what happens to STO, heated to 1800 K \cite{Ligny1996}, a remarkably high temperature, broadly compatible with the highest energy scale in the phonon spectrum of these solids ($\approx$ 100 meV) \cite{Rushchanskii2012,Aschauer2014}. A systematic difference in the specific heat evolves at lower temperatures. As is shown in the inset of Fig.~\ref{Cp_QP}(a), all three solids show a peak in the temperature dependence of $\frac{C}{T^3}$. In the case of STO, this peak is known to be caused by the presence of Einstein modes \cite{BURNS1980811,McCalla2019}. Similar peaks are visible for KTO and for ETO, and the position of this peak shifts with the increase in the molar mass. In STO (184 g/mol), it occurs at 30 K, in ETO (248 g/mol) at 25 K, and in KTO (267 g/mol) at 12 K. In the case of ETO, a distinct additional contribution shoots up below 15 K, which signals strong magnetic fluctuations above the N\'eel ordering temperature of the Eu$^{2+}$ spins. 

\subsection{Field dependence }
As reported previously \cite{Petrovic2013}, the specific heat in ETO displays a strong dependence on magnetic field. In order to separate the magnetic $C_{mag}$ and the phononic  $C_{phon}$ contribution, we subtracted from the total specific heat of EuTiO$_3$ the measured specific heat of SrTiO$_3$ (with a temperature re-scaled by a factor of 0.83 such that the positions of the $C/T^3$ maxima of both materials match). Note that this scaling factor is close to the ratio of the molar masses $\sqrt{M_\mathrm{STO}/M_\mathrm{ETO}}=0.86$,  the expected ratio of their Debye temperatures. Figure~\ref{Cp_QP}(b) shows the magnetic heat capacity $C_{mag}/T$ of EuTiO$_3$ for different magnetic fields from 0 up to $10\,$T applied along a $[100]_\mathrm{c}$ direction of the cubic room temperature phase. The sharp zero-field anomaly signals the antiferromagnetic order at $T_\mathrm{N}=5.5\,$K that gets strongly broadened already in a field of $1\,$T. This reflects the magnetic-field-induced switching to the polarized magnetic state, which sets in around 1.2~T in ETO (see, e.g., the supplement \cite{SM} or \cite{Midya2016}) and, consequently, the $C_{mag}/T$ data in larger fields no longer signal a spontaneous magnetic ordering transition, but a continuous evolution of magnetic entropy as it is the case for ferromagnets in an external magnetic field. In fact, the $C_{mag}/T$ data of EuTiO$_3$ strongly resembles the corresponding data obtained on the Eu$^{2+}$-based ferromagnet EuC$_2$, which orders at $T_\mathrm{C}=14$~K \cite{Heyer_2011}. The magnetic entropy obtained by integrating $C_{mag}/T$ is displayed in the inset of Fig.~\ref{Cp_QP}(b) and reveals that the full magnetic entropy $N_A k_B \ln(2S+1)$ of a spin 7/2 system is reached above about $15\,$K for zero field and also for 1 T, whereas for fields above 2~T the entropy evolution drastically broadens and extends to much higher temperatures.

\begin{figure}
\begin{center}
\centering
\includegraphics[width=0.48\textwidth]{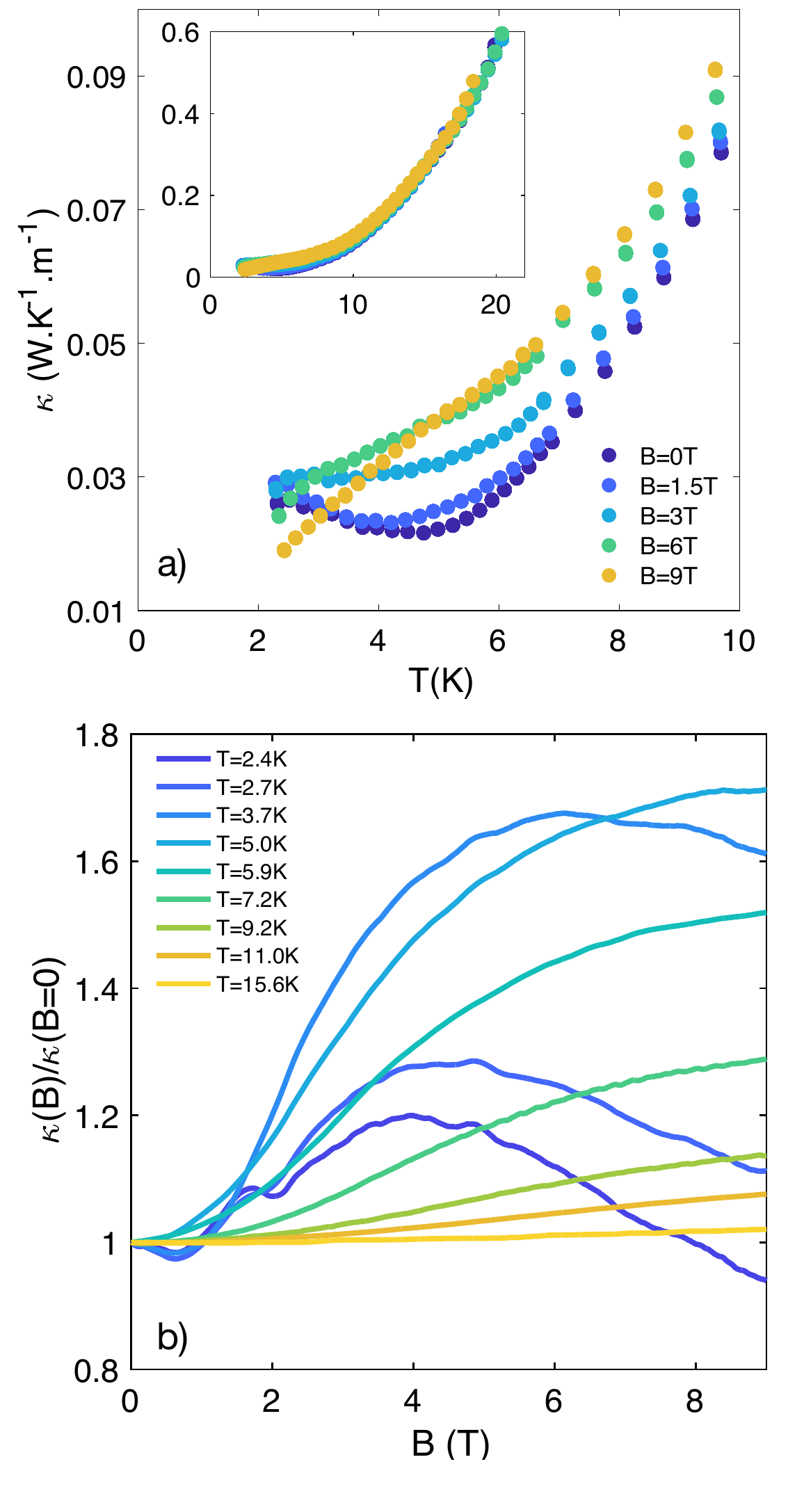}
\caption{{\bf{Field dependence of thermal conductivity in EuTiO$_3$} } a) $\kappa$ vs. T from 2 K to 10 K for B = 0 T to 9 T. b) Normalised $\kappa$ vs. B for T = 2.4 K to 15.6 K for B$//$[001].}
\label{Fig_kappa_H}
\end{center}
\end{figure}

The  field dependence of $\kappa$, shown in Fig. \ref{Fig_kappa_H},  further points to an intricate coupling between Eu$^{2+}$ spins and heat-carrying phonons. Thermal conductivity begins to depend on magnetic field below 15 K. Interestingly, as seen  in the inset of Fig. \ref{Cp_QP}, this is the temperature below which there is a significant magnetic entropy.  

Above $T_N$,  magnetic field slightly amplifies $\kappa$, indicating a weakening of the spin-induced localisation of phonons. 
%At 2 K, magnetization  is linear up to 1.2 T and then saturates to 7$\mu_B$ per formula unit, implying that all the Eu$^{2+}$ spins become aligned above 2 T \cite{Midya2016}. The additional field scales in $\kappa$ imply that the energy levels of the magnetic ions and their coupling to phonon modes play a role in setting the amplitude of thermal conductivity.
This field-induced amplification  of  $\kappa$ in ETO is modest, in contrast with to  the thirty-fold field-induced amplification of the ultra-low thermal conductivity in Tb$_2$Ti$_2$O$_7$ \cite{Li2013}. 

Below 3~K, well below the N\'eel temperature,   the field-dependence becomes non-monotonic (See Fig. \ref{Fig_kappa_H}. We leave the quantitative understanding of the field dependence of $\kappa$ in EuTiO$_3$ as a subject of study for future investigations.

\subsection{Thermal diffusivity and phonon mean-free-path}
\begin{figure}
\begin{center}
\centering
\includegraphics[width=0.5\textwidth]{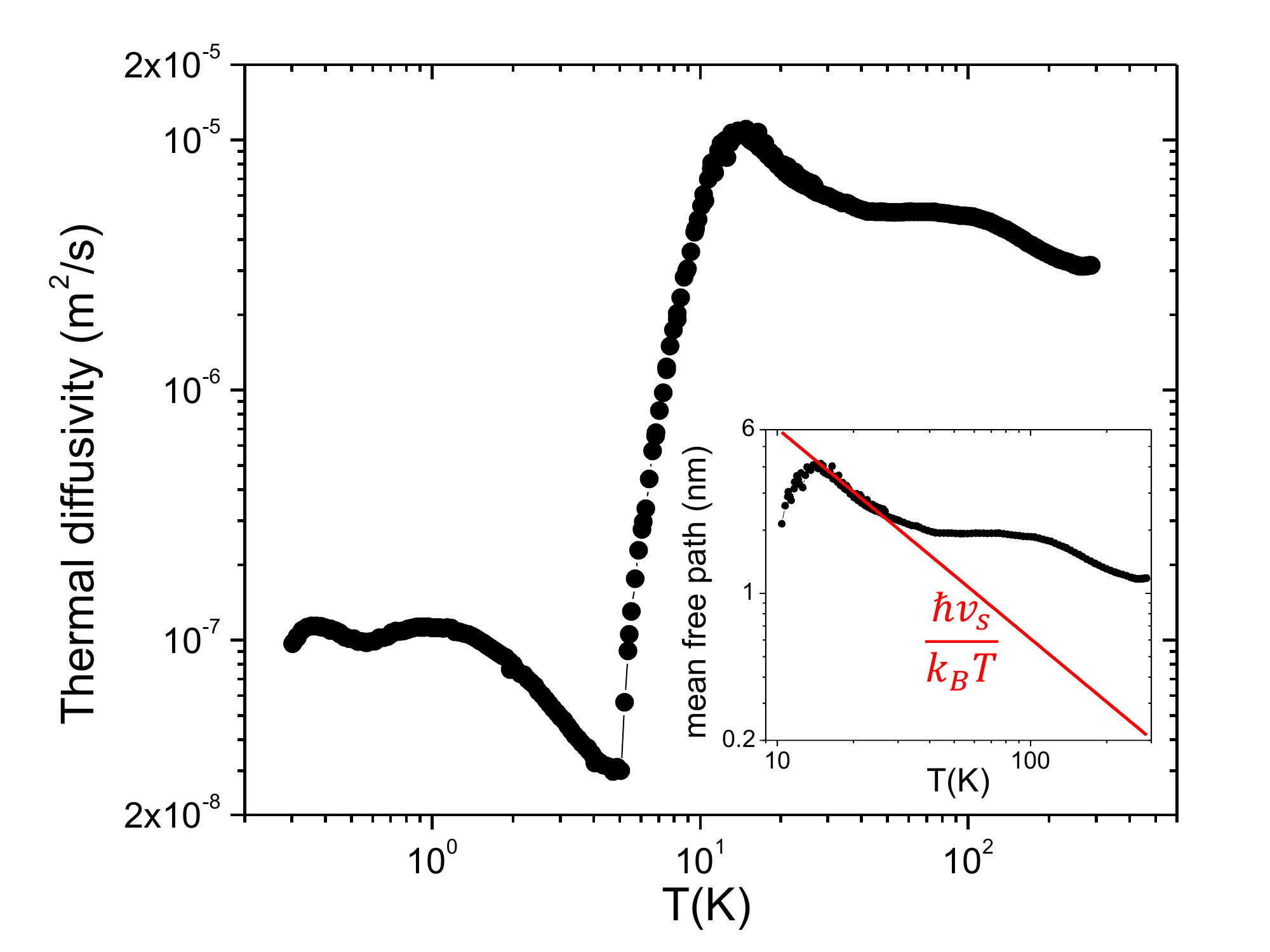}
\caption{{\bf{Thermal Diffusivity of EuTiO$_3$ and the mean-free-path of phonons:}} Temperature dependence of thermal diffusivity, $D$, by taking the ratio of thermal conductivity and the specific heat per volume in sample C1. Note the drastic drop across the magnetic transition. The insets shows the Temperature dependence of the mean free path in the paramagnetic phase, assuming that all phonons have the same scattering time and velocity for the same sample. The red solid line represents the inverse of the wavevector of the thermally excited acoustic phonons.}
\label{Fig-diffusivity}
\end{center}
\end{figure}
Replacing Sr by Eu reduces the thermal conductivity in a wide temperature range and enhances the specific heat below 15 K. Therefore, the ratio of thermal conductivity  (in W/(K.m)) to specific heat per volume (in J/(K. m$^3$)), i.e. the thermal diffusivity, $D$, (in m$^2$/s) is drastically modified. It is plotted in Fig. \ref{Fig-diffusivity}. The most striking feature of $D (T)$ is its two-orders-of-magnitude drop at the N\'eel temperature. Within the entry to the antiferromagnetically ordered phase, at 5.5~K, it becomes exceptionally low. Its minimum, 0.03 mm$^2$/s, is almost two orders of magnitude lower than the thermal diffusivity of a typical glass \cite{Yang1992}.  
This low thermal diffusivity, a consequence of the combination of an unusually low lattice thermal conductivity and a very large magnetic entropy may find applications in heat management in a cryogenic context. 

The thermal conductivity of an insulator can be written as:
\begin{equation}
\kappa=\sum_{s,\bf{q}} C_{s}(\bf{q})v^2_{s}(\bf{q}) \tau_{s}(\bf{q}) 
\label{Eq1}
\end{equation}
The index $s$ refers to different modes and $\bf{q}$ is the wave-vector. $C_{s}$, $v_{s}$ and  $\tau_{s}$ are, specific heat, velocity and scattering time. There are modes contributing to the total specific heat ($C=\sum_s C_s$ ), but not to thermal conductivity, because of their negligible velocity.  Theoretically \cite{Bussmann_Holder_2012}, paramagnons  in the paramagnetic state  have no dispersion and therefore do not carry heat. They can, however, reduce the phonon thermal transport. In ETO, phonons not only dominate thermal conductivity, but also, at least down to 15 K, the specific heat. Therefore, one can simply write: $\kappa=1/3C v_s\ell_{ph}$. This neglects the $q$ dependence of the scattering time and assumes that all modes have the same velocity. Taking  $v_s$= 6.8 km/s,  the measured sound velocity in STO \cite{Okai1975}, and in reasonable agreement with the dispersion of acoustic branches in ETO \cite{Rushchanskii2012}, one can estimate 
$\ell_{ph}$, shown in  the inset of  Fig. \ref{Fig-diffusivity}. Comparing it to the inverse of the wave-vector of thermally excited phonons: $q_s=\frac{k_BT}{\hbar v_s}$, one finds that below 20 K, $q_s\ell_{ph} \cong 1 $ , reminiscent of the Anderson localization. There is a drop at 15K, below which specific heat is no more phonon dominated.

\section{Discussion}
Evidence for a coupling between spin and lattice degrees of freedom in this compound was first reported by Katsufuji and Takagi \cite{Katsufuji2001}, who found that when the spins order at 5.5 K, the electric permittivity of ETO drops by 7 percent and this drop is suppressed by the application of a magnetic field of the order of 3 T. This magneto-electric effect implies coupling between Eu$^{2+}$ spins and the soft mode governing the electric permittivity. Reuvekamp {\it{et al.}} \cite{Reuvekamp_2015} have found a quantitative agreement between the amplitude of the magneto-electric effect and the low-temperature magnetostriction of the system. 

Our main finding is that lattice-spin coupling drastically attenuates the propagation of heat in ETO, even at temperatures where magnetic ordering  is absent and magnetic entropy is practically saturated at its maximum value $k_B\ln(2S+1)$/spin. This implies that even when the spins are randomly oriented, heat-carrying phonons couple to Eu$^{2+}$ states and their large magnetic moments (6.9-7 $\mu_B$). According to ref.~\cite{Malyi2022}, without incorporating the loss of spin symmetry, the DFT calculations cannot explain the absence of metallicity and the finite band gap of the paramagnetic phase. 
\begin{figure}
\begin{center}
\centering
\includegraphics[width=0.48\textwidth]{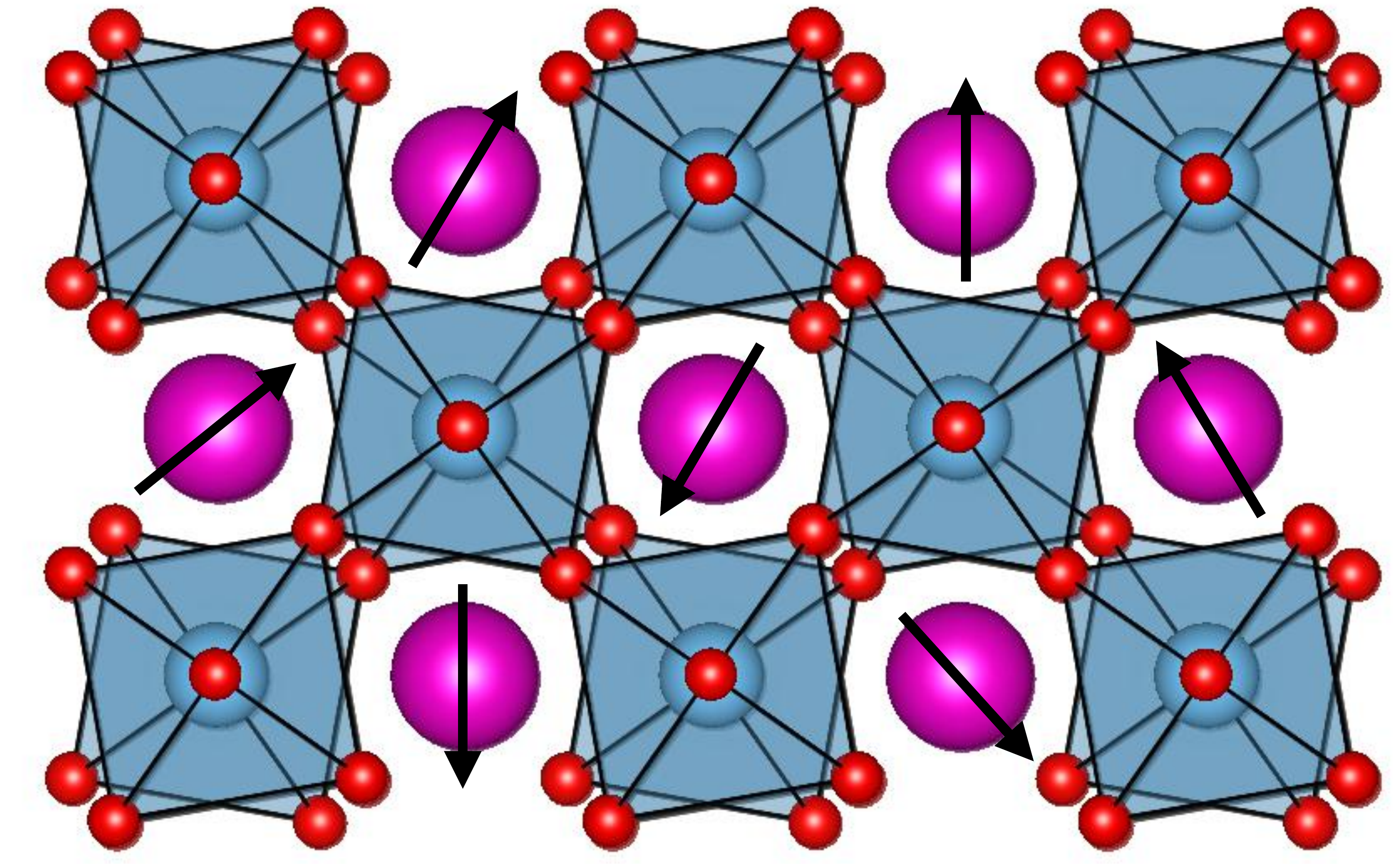}
\caption{{\bf{Crystal structure of EuTiO$_3$ in its paramagnetic phase:}} Random magnetic moments at Eu sites are located between TiO$_6$ octahedra, which are tilted off each other. Superexchange interaction between these spins involves Ti and oxygen atoms. This can impede phonons to have a well-defined wave-vector over long distances. }
\label{Fig-ST}
\end{center}
\end{figure}

The random orientation of magnetic moments at Eu sites (See Fig. \ref{Fig-ST}) is the most plausible source of phonon localization. The superexchange interaction between Eu spins occurs through Ti ions \cite{Akamatsu2011}. The inter-atomic force constant can depend on the relative orientation of spins. The calculated phonon frequencies for parallel or anti-parallel alignment of adjacent spins are not the same \cite{Fennie2006}, which implies that phonons cannot keep a well-defined dispersion in presence of random spin orientation. 

The narrow energy separation between the Eu$^{2+}$ energy level and the chemical potential may also play a role. In some Eu-based metals, the thermoelectric response has been  linked to the temperature dependence of Eu valence \cite{Stockert2020}. Remarkably, a theoretical study \cite{Feng_2015} has concluded that the contribution of optical phonons to the overall lattice thermal conductivity is unusually large in this lattice structure.  A coupling between heat-carrying optical phonons and Eu valence may be the source of observed $\kappa$ attenuation. 

Inelastic neutron scattering is a promising probe of this physics. In the case of Tb$_3$Ga$_5$O$_{12}$, for example, it documented the coupling between spin and lattice \cite{Petit2021}). A recent study on STO \cite{Fauque2022} has found evidence for an unusual hybridization between acoustic and optic phonon branches. No neutron scattering data is presently available to compare ETO with STO.

%For Eu$^{2+}$(4$f^{7}$), $J$=$S$=$\frac{7}{2}$ ($L$ = 0), the first expected resonance between $J_z$=$\pm\frac{7}{2}$ states at  $T$ = 2.4 K is $B$ = 1 T, which is  close to the first observe minimum. Other resonances can occur within the other splits levels of the ground state such as at 2.1 T, that can be attributed to a resonance condition between the $\frac{\pm3}{2}$ levels, expected at 2.4 T), but also with the lower branch of the first-excitation level that could lead to another minimum above 10 T. If so, the first-excitation level and the ground state would be separated to about few meV and could give rise to a resonant scattering even at zero magnetic field.

%In the picture drawn above, this happens because the introduction of magnetic field attenuates the randomness of the spin orientations, which is the driver of the low thermal conductivity. 
%In the $\alpha$ phase (below 5 K), the evolution with magnetic field is more prominent and non-monotonous. $\kappa$ peaks at a temperature-dependent magnetic field of 4-6 T. 

%What attenuates heat transport in the $\alpha-$ phase is therefore a strong, but indirect coupling between acoustic phonons and localized spins.

Finally, let us note that the glass-like thermal conductivity of EuTiO$_3$, in contrast to spin-liquid crystals, occurs in a solid with a simple G-type anti-ferromagnetic ground state \cite{McGuire}. A formal theoretical  treatment may be achieved by complementing the picture drawn by Eq.\ref{Eq1} with an `off-diagonal' coupling between different vibrational states \cite{Simoncelli2019,Isaeva2019,HANUS2021100344}:
\begin{equation}
\kappa_{od}=\sum_{ss',\bf{q}}^{s\neq s'} C_{ss'}(\bf{q})v^2_{ss'}(\bf{q}) \tau_{ss'}(\bf{q}) 
\label{Eq2}
\end{equation}

This equation was conceived for non-magnetic crystals, in which harmonic coupling occurs across phonon branches \cite{Simoncelli2019,HANUS2021100344}. It can be extended to magnetic modes.

\section{Acknowledgements}
We thank Annette Bussman-Holder, Yo Machida, Igor Mazin and Nicola Spaldin for stimulating discussions. This work was supported in France by JEIP-Coll\`{e}ge de France, and by the Agence Nationale de la Recherche (ANR-18-CE92-0020-01; ANR-19-CE30-0014-04) and a grant  by the Ile de France
regional council. In Germany, it was supported by the DFG (German Research Foundation) via Project No. LO 818/6-1. I.H.I.In Japna, it was supported by the Japan Society for the Promotion of Science (JSPS) KAKENHI Grant No. 18H03686 and the Japan Science and Technology Agency (JST) CREST Grant No. JPMJCR19K2.

\bibliography{KappaETO}

\clearpage
% Add 'S' to the numbering inside the supplement
\renewcommand{\thesection}{S\arabic{section}}
\renewcommand{\thetable}{S\arabic{table}}
\renewcommand{\thefigure}{S\arabic{figure}}
\renewcommand{\theequation}{S\arabic{equation}}
\setcounter{section}{0}
\setcounter{figure}{0}
\setcounter{table}{0}
\setcounter{equation}{0}
%\setcounter{thebibliography}{0}

% Produces the bibliography via BibTeX.
\begin{widetext}
%\newpage

\begin{center}{\large\bf Supplementary Materials for \emph{Glass-like thermal conductivity and narrow insulating gap of EuTiO$_3$ }}\\
\end{center}

\section{Crystal growth }

EuTiO$_3$ single crystals were grown in two different labs.
In Cologne, the floating-zone technique was used as described in Ref. \cite{Engelmeyer2019}. In order to prevent the formation of the Eu$_2$Ti$_2$O$_7$ pyrochlore phase, the growth was performed in Argon atmosphere and without any preliminary powder reaction, similar to Ref. \cite{Katsufuji1999}. Powders of Eu$_2$O$_3$, TiO, and TiO$_2$ were mixed in a ratio to reach EuTiO$_{3-\delta}$, pressed to a rod, and installed into the floating-zone furnace. In a series of growth attempts, it turned out that it was necessary to mix the above materials with a nominal oxygen deficiency of $\delta =0.05$.  This allowed  to compensate possible off-stoichiometries of the starting materials or an oxygen capture during the growth process. Single crystallinity of the grown crystal was confirmed by Laue images and phase purity was verified by X-ray powder diffraction.

In Tsukuba, similarly to the above description, EuTiO$_3$ crystals were prepared by the floating zone method \cite{Tomioka2018}.

\section{Magnetization}
\label{mag_pristine}

Figure~\ref{fig_mag}(a) displays magnetization data M(T) of a EuTiO$_3$ single crystal for different magnetic fields between 10 and 500~mT applied along a $[100]_\mathrm{c}$ direction of the cubic room temperature phase. At the lowest fields, the antiferromagnetic ordering causes a maximum in $M(T)$. This peak becomes a kink at an intermediate field of 500~mT. The corresponding inverse susceptibility follows a straight line over the entire temperature range from 300~K down to about 7~K (See panels~(b) and ~(c)). The Curie-Weiss fit yields an effective magnet moment of $\mu_{eff}=7.8\,\mu_\mathrm{B}$/f.u. and a positive Weiss temperature of $\theta_\mathrm{W}=3.55\,$K, which signals a net ferromagnetic coupling. These values agree with the literature. The positive $\theta_\mathrm{W}$ has been explained by the fact that in EuTiO$_3$ the antiferromagnetic coupling of the Eu$^{2+}$ spins to 6 nearest neighbors (NN) is overcompensated by a larger ferromagnetic coupling to 12 next nearest neighbors\cite{McGuire,Midya2016,Chien1974,Akamatsu2012,Petrovic2013}.  
\begin{figure}[h]
	\centering
	\includegraphics[height=75mm]{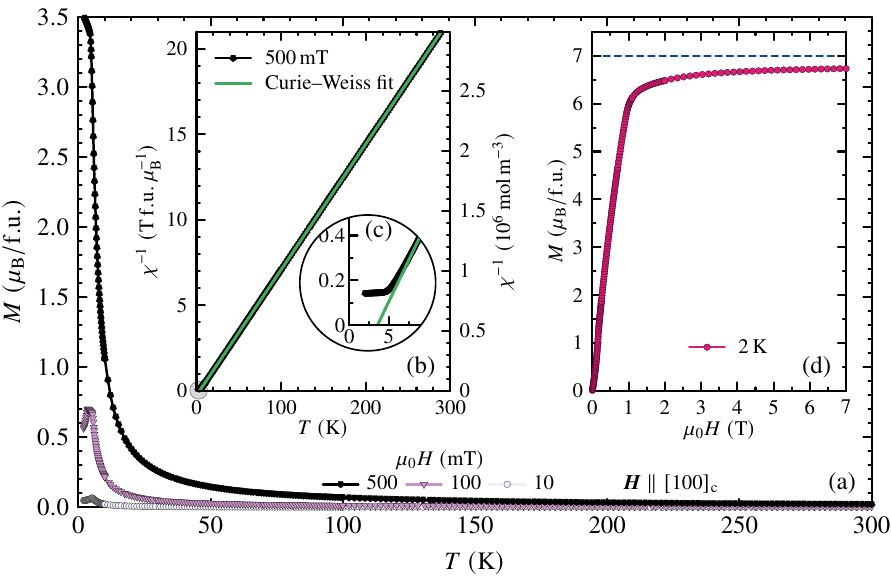}
    \includegraphics[height=73mm]{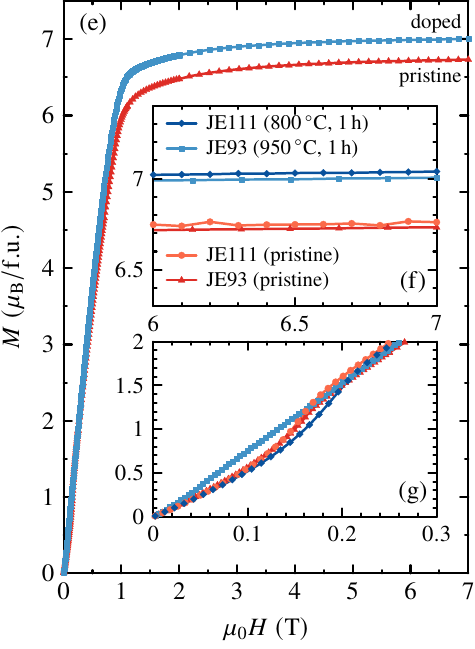} 
	\caption{(a)~Magnetization of EuTiO$_3$ as a function of temperature for different magnetic fields applied along the cubic $[100]_\mathrm{c}$ direction. (b)~Inverse susceptibility $\chi^{-1}=\mu_0H/M$ at 500~mT with a Curie--Weiss fit (line) for $T > 20$~K (c)~Detailed view of the low-temperature regime, revealing the Weiss temperature $\theta_\mathrm{W}=3.55$~K. (d)~Magnetization curve measured at 2~K, which almost reaches the saturation magnetization expected for a spin 7/2 system (dashed line) (e)~Magnetization curves measured on 2 different  pristine EuTiO$_3$ single crystals (JE93, JE111), which both were insulating, in comparison to 2 metallic EuTiO$_3$ samples, which were weakly electron doped by vacuum annealing~\cite{Engelmeyer2019}. Insets show enlarged views of the same data in the region of (f) the saturation magnetization and of (g) a spin-flop transition occurring at $\approx 0.2\,$T when the field is applied along a cubic [100]$_c$ direction (3 samples), but remains absent for $H [110]_c$ (doped sample JE93).} 
	\label{fig_mag}
\end{figure}

Note that there is no magnetic frustration resulting from the ferromagnetic and antiferromagnetic couplings because both couplings support the G-type antiferromagnetic order. Nevertheless, a comparatively weak magnetic field of about 1.2 T is sufficient to induce a polarized magnetic state. This is because only the weaker NN antiferromagnetic coupling needs to be overcome. As shown  in Fig.~\ref{fig_mag}(d), the saturation magnetization $M_\mathrm{sat}\simeq 6.73\,\mu_\mathrm{B}$/f.u. reaches about 96\,\% of the expected spin-only value of $M_\mathrm{sat}\simeq g\mu_\mathrm{B} S_z =7\mu_\mathrm{B}$ for Eu$^{2+}$ with half-filled $4f$ shell. An analogous 4\,\% reduction is obtained from the Curie-Weiss fit in the paramagnetic phase for the Curie constant, which results in a 2\,\% reduction of the effective magnetic moment $\mu_{eff}=7.8\,\mu_\mathrm{B}$/f.u. in comparison to the expected $\mu_{eff}=g\sqrt{S(S+1)}\mu_\mathrm{B}=7.94\,\mu_\mathrm{B}$ for $S=7/2$.

As is shown in Fig~\ref{fig_mag}(e) this weakly reduced saturation magnetization has been observed in various pristine insulating samples that were obtained from different growth attempts. In contrast, however, other EuTiO$_3$ samples that were cut from the same mother crystals, but were made conducting via vacuum annealing at high temperature, see e.g.~\cite{Engelmeyer2019}, reached the expected saturation magnetization of $7\mu_\mathrm{B}$/Eu$^{2+}$. As it is known that the vacuum annealing of STO and ETO causes electron doping which is traced back to a reduction of the oxygen content, the different saturation magnetizations of pristine versus doped ETO can be interpreted as follows:  appearently, the pristine ETO samples have a weak oxygen surplus that results in a corresponding partial oxidization of Eu$^{2+}$ to Eu$^{3+}$ which has a  vanishing magnetic moment thereby reducing the overall magnetization. Vice versa, with decreasing oxygen content, the magnetization of pristine ETO should first increase but remain insulating until all Eu$^{3+}$ ions are reduced to Eu$^{2+}$ and only with further oxygen reduction a metal-insulator transition sets in via electron doping into the 3d band of Ti~\cite{Engelmeyer2019}. 

\section{Methods for measuring thermal conductivity}
We measured thermal conductivity with a standard one-heater-two-thermometers method. Above 2K, the measurements were carried out with a home-built setup in a Quantum Design PPMS. Down to 100mK, the measurements were realized in a Dilution refrigerator system. The sensors used for the measurement of ETO were Cernox chips Cx-1030 and RuO$_2$ thermometers. Above 20 K and up to room temperature, type E thermocouples were used. Close attention was paid to the overlap between different sets of experimental data. The thermometers were directly glued to the samples with Dupont 4922N silver paste. These sensors were also thermally isolated from the copper sample holder by long manganin wires with very low thermal conductance. The ETO/STO samples were connected to a heat sink (copper finger) with Dupont 4922N silver paste on one side and to a RuO$_2$-resistor (used as a heater) on the other side. A heating current was applied by flowing a current $I$ through the heater from a DC current source (Keithley 6220). We evaluated the heating power from the relation $I \times V$ with $V$: the electric voltage measured across the heater by a digital
multimeter (Keithley 2000). $\kappa$ was controlled to be independent of the applied thermal gradient $\Delta T$ by changing $\Delta T/T$. $(\Delta T/T)_{max}$ was kept under $10\%$. Finally, calibration of the sensors was realized during each experiment and upon verification.

\section{Sample dependence}
\begin{figure}
\begin{center}
\centering
\includegraphics[width=0.9\textwidth]{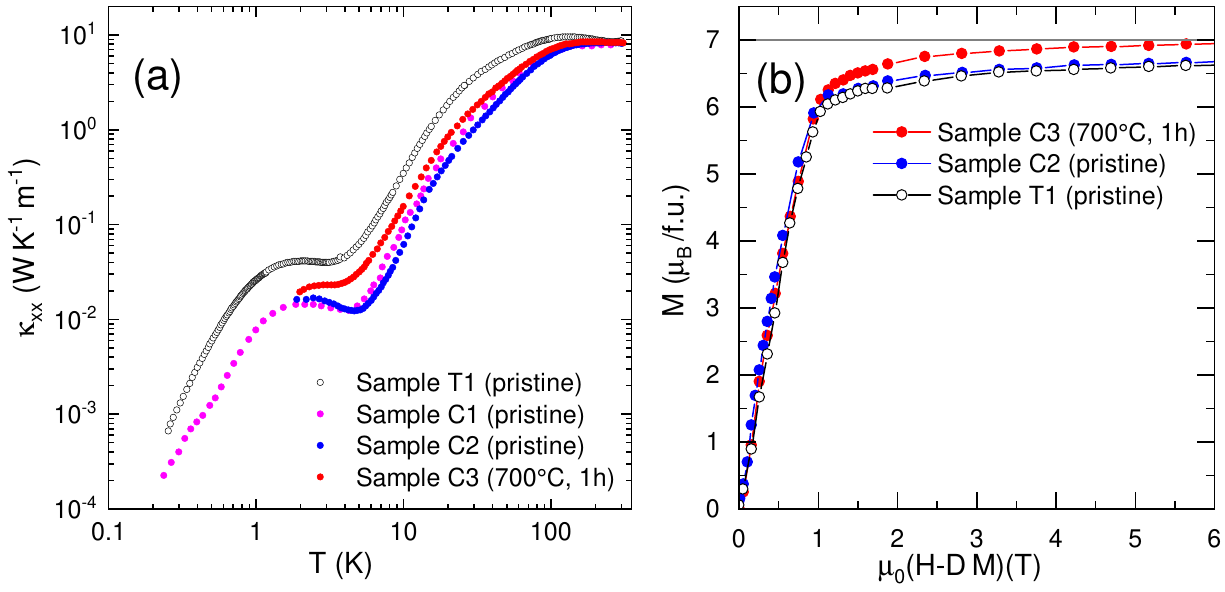}
\caption{(a) Comparison of thermal conductivity data $\kappa(T)$ of EuTiO$_3$, which were  measured on different samples, which stem from different sources (T1 versus C1, C2, C3) (b) Magnetization data measured on the same samples apart from C1, that got broken in the meanwhile. Note that the field axis was corrected for demagnetization effects to account  for the rather different sample geometries.}
\label{Fig-SM1}
\end{center}
\end{figure}
\color{black}

As discussed above, the magnetization data of pristine ETO crystals indicate the presence of a certain mixture of Eu$^{2+}$ and Eu$^{3+}$ valencies, and it is known from many intermetallic materials that Eu (or other rare-earth) valence instabilities can drastically influence the magnetic and/or transport properties~\cite{Stockert2020}. Therefore, we studied several EuTiO$_3$ crystals in order to check whether there is some correlation between the surprisingly low, glass-like thermal conductivity and the reduced saturation magnetization. As can be seen from Fig.~\ref{Fig-SM1}(a), the overall variation of the $\kappa(T)$ curves obtained on the different samples is covered by the data obtained on samples C1 and T1, which are discussed in the main article. The magnetization data of the pristine samples C2 and T1 are essentially identical to each other and agree with the data of the other pristine samples shown in Fig.\ref{fig_mag}. While this is not surprising for C1, which is another piece from the same mother crystal JE111, this was not necessarily to be expected for T1, which was grown independently under somewhat different conditions (see above) and it is evident from panel (a) that overall the $\kappa(T)$ curve  of T1 is lying above those measured on the samples C1, C2, and C3. On the other hand, the magnetization of C3 is the only one which essentially reaches $7\mu_\mathrm{B}$. This can be traced back to the fact this sample was vacuum annealed for 1~h at 700$^\circ$C. Resistivity measurements before and after annealing did not show a  measurable difference, whereas the enhanced magnetization indicates a significant variation of Eu valence towards almost pure Eu$^{2+}$. Nevertheless, the thermal conductivity of this particular C3 sample remains as low and glassy as those of all the ETO samples. Thus, we conclude that the weak Eu$^{2+/3+}$ off-stoichiometry is not the main mechanism that induces the unusual glass-like phonon heat transport in the ETO crystals. 
\begin{figure}
\begin{center}
\centering
\includegraphics[width=0.9\textwidth]{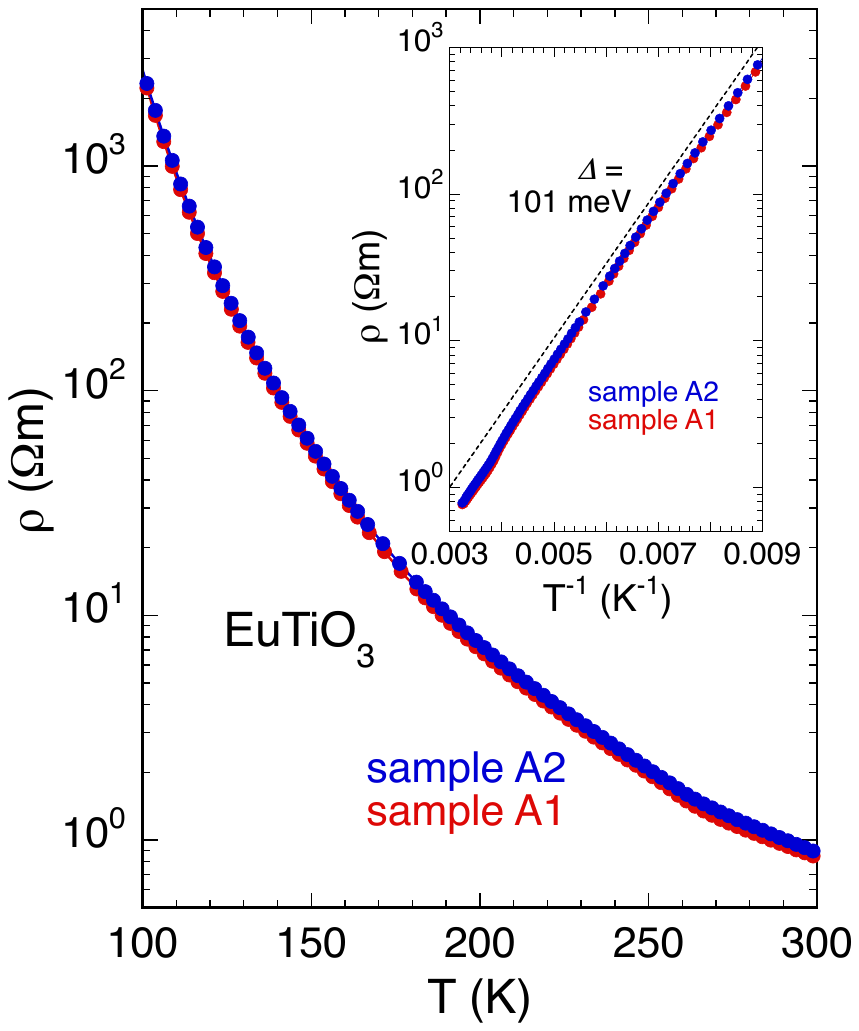}
\caption{Temperature profiles of resistivity in two EuTiO$_3$ crystals (samples A1 and A2). Inset displays ln $\rho$ \textit{vs.} T$^{-1}$. Dotted line is calculated using D = 101 meV.}
\label{Fig-reprod-rho}
\end{center}
\end{figure}
\color{black}

We also confirmed the activated behavior of the resistivity in several samples measured in three different laboratories. Figure \ref{Fig-reprod-rho} indicates temperature profiles of resistivity, $\rho$ of two EuTiO$_3$ crystals prepared by the floating zone method \cite{Tomioka2018}. Tiny anomalies can be discerned at near 260 K, which is the transition between cubic and tetragonal structure, consistent with the discussion in Sec. II. A of the main paper. The inset  shows ln$\rho$  \textit{vs.} T$^{-1}$ of these crystals. The estimated value of the activation gap, $\Delta$ is comparable to that of samples T1 and T2 in Figure 1(b).
\section{The Hall coefficient}

\begin{figure}
\begin{center}
\centering
\includegraphics[width=0.9\textwidth]{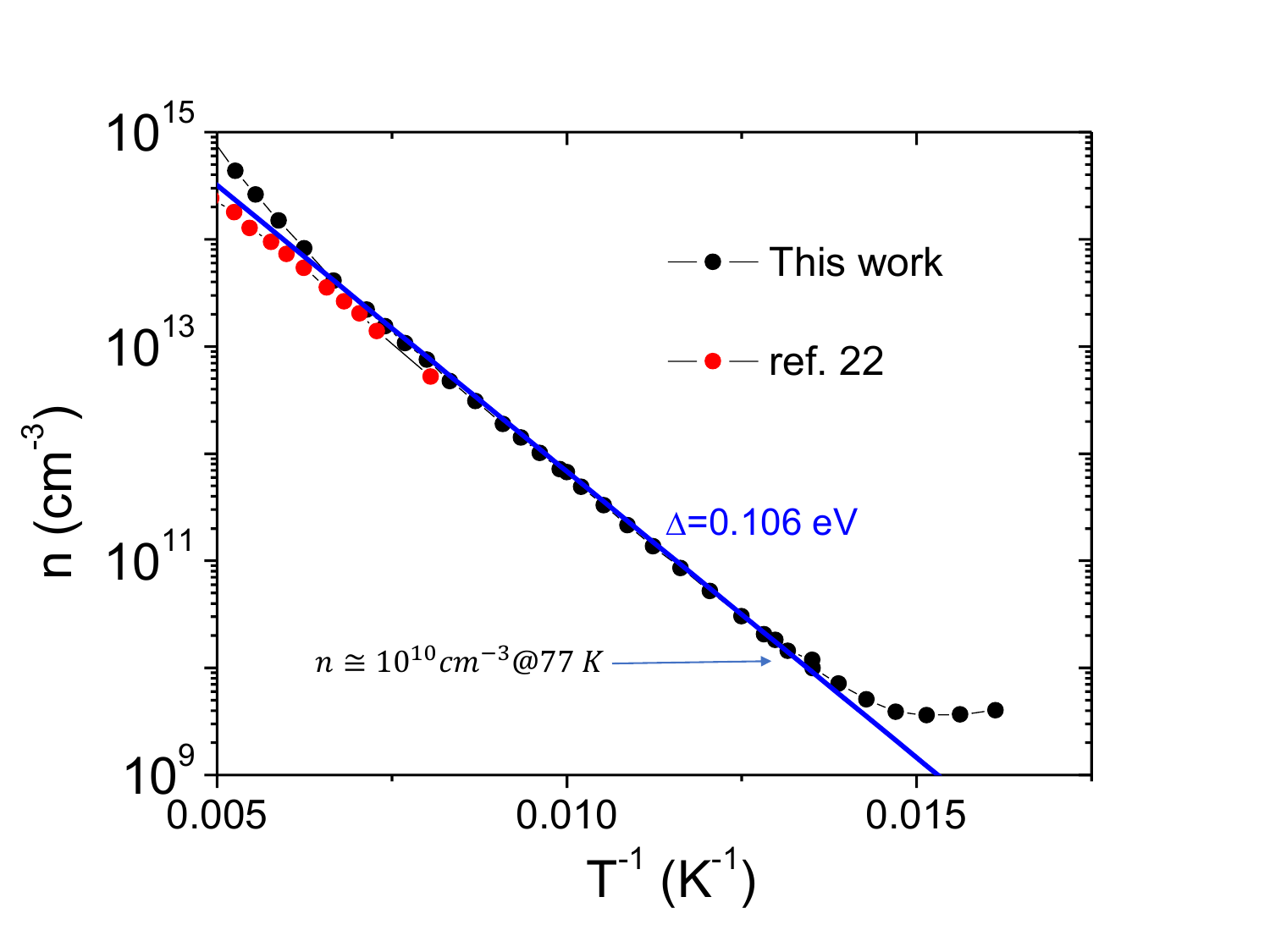}
\caption{The Hall carrier density as a function of the inverse of a temperature in an Arrhenius plot. }
\label{Fig-Hall}
\end{center}
\end{figure}
\color{black}
As shown in Fig.\ref{Fig-Hall}, we found that the Hall carrier density displays an Arrhenius behavior from 180 K down to 70 K. A deviation starts below the latter temperature, when the resistance was in the range of $10^8$ ohms rendering the data unreliable. The corresponding carrier density was   $10^{10}$cm $^{-3}$. At very low temperature, the saturated Hall number corresponds to  a carrier density of $\approx 10^9$cm $^{-3}$. 

\section{Hysteresis near the structural transition}
The structural transition in ETO is believed to be similar to the structural transition in STO an antiferrodistortive transition in which adjacent octahedra rotate in opposite orientations \cite{Bussmann_Holder_2012}. In the case of STO, this transition is known as a rare case of second order structural phase transition. However, in our study of resistivity, we found that the anomaly at $\approx$ 260 K presents a hysteresis, which was found to be reproducible. The data is shown in Fig \ref{hysteresis}. This may point to the first-order nature of this structural transition in ETO. 
\begin{figure}
\begin{center}
\centering
\includegraphics[width=0.9\textwidth]{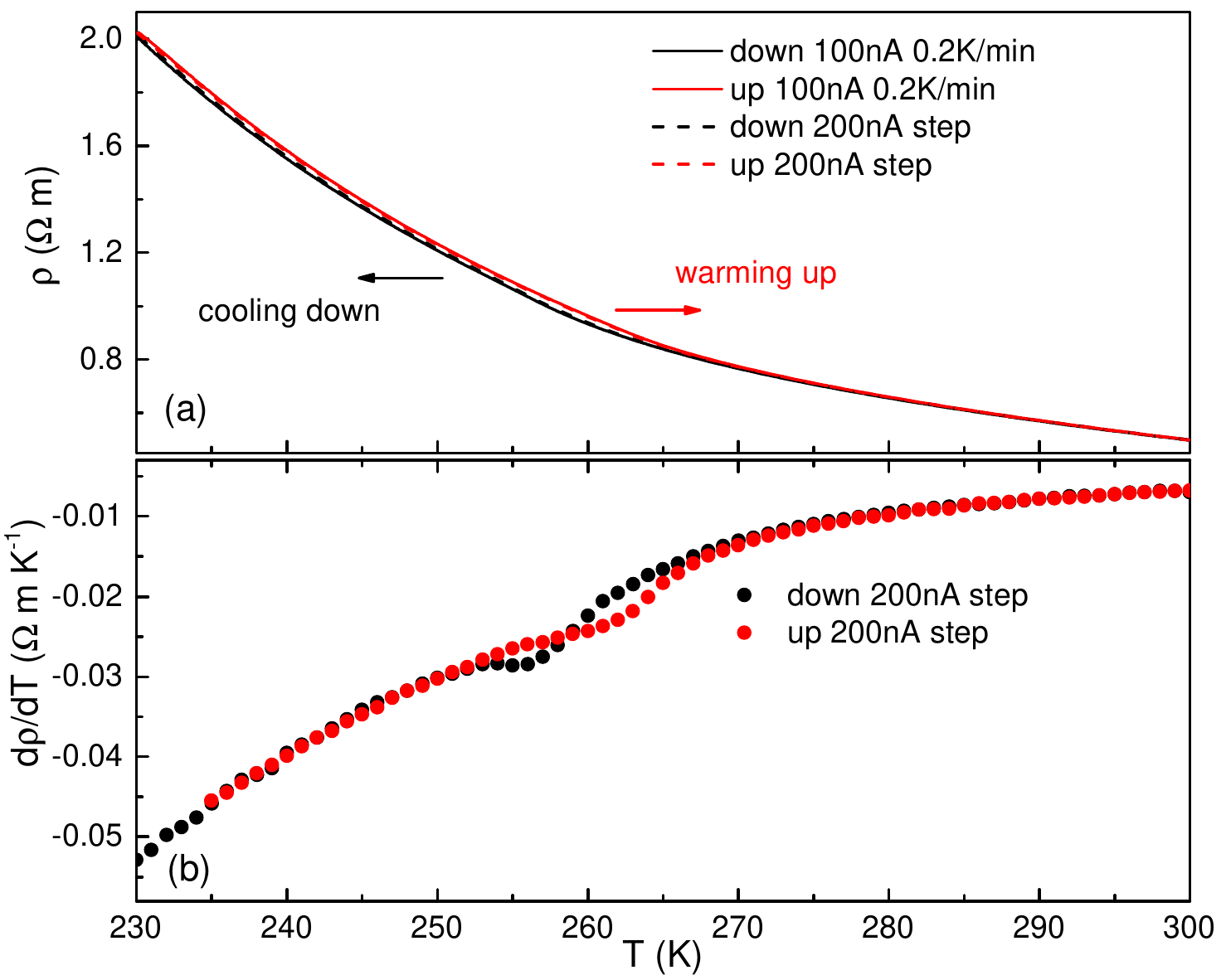}
\caption{a) Temperature dependence of resistivity near the structural transition for upward and downward sweeps either with a continuous ramp or by steps. b) The temperature derivative of resistivity for an upward and a downward sweeps with discrete steps. }
\label{hysteresis}
\end{center}
\end{figure}
\end{widetext}
\end{document}